\documentclass[11pt,epsf,letterpaper]{article}%
\usepackage{setspace}
\usepackage{color}
\usepackage{graphicx}
\usepackage{amsmath}
\usepackage{amsfonts}
\usepackage{verbatim}
\usepackage{amssymb}
\usepackage[normalem]{ulem}
\usepackage{float}

\usepackage{url}

\usepackage[title]{appendix}

\providecommand{\U}[1]{\protect\rule{.1in}{.1in}}
\newcommand{\be}{\begin{equation}}
\newcommand{\ee}{\end{equation}}
\newcommand{\ba}{\begin{eqnarray}}
\newcommand{\ea}{\end{eqnarray}}

\begin{document}

\title{{\bf Some insight into Feynman's approach to electromagnetism}}

\author{Marco Di Mauro$^{1,2,\clubsuit}$, Roberto De
Luca$^{1,\spadesuit}$, Salvatore Esposito$^{2,\diamondsuit}$,
Adele Naddeo$^{2,3,\heartsuit}$ \\
$^{1}$\footnotesize{Dipartimento di Fisica ``E.R. Caianiello",
Universit\'a di Salerno,
Via Giovanni Paolo II, 84084 Fisciano, Italy.}\\
$^{2}$\footnotesize{INFN Sezione di Napoli, Via Cinthia, 80126
Naples, Italy.}\\
$^{3}$E-mail: anaddeo@na.infn.it\\
ORCIDs: $^{\clubsuit}$0000-0002-1532-8274;
$^\spadesuit$0000-0001-8137-6904\\
$^\diamondsuit$0000-0003-3099-5574;
$^\heartsuit$0000-0002-0723-9343.}\maketitle

\begin{abstract}
\noindent We retrace an \emph{ab initio} relativistic derivation
of Maxwell's equations that was developed by Feynman in
unpublished notes, clarifying the analogies and the differences
with analogous treatments present in the literature. Unlike the
latter, Feynman's approach stands out because it considers
electromagnetic potentials as primary, reflecting his ideas about
the quantum foundations of electromagnetism. Some considerations
about the foundations of special relativity, which are naturally
suggested by this approach, are given in appendix.
\end{abstract}


\section{Introduction}

In previous work \cite{DeLuca:2019ija}, we have begun the study of
an unpublished original approach to electromagnetism, which was
developed by Feynman for teaching purposes. This approach was
retraced to two sets of unpublished notes. The first part is
contained in some handwritten pages \cite{Gottlieb}, dating back
to 1963. The second part has been located in some lectures given
by Feynman at the Hughes Aircraft Company in the end of the '60s
\cite{FeynmanHughes}. In the handwritten pages, Feynman traces an
outline of development of electromagnetism from scratch,
originally intended to replace the treatment developed in
Feynman's undergraduate lectures at Caltech
\cite{Feynman:1963uxa}. He claims his intention to start from
relativistic invariance, and to assign a privileged role to
electromagnetic potentials, rather than to fields. On that
occasion, he does not develop the proposed outline, but limits
himself to a derivation of the Lorentz force from the empirically
based assumption of relativistic invariance of electric charge.
This derivation is actually very original, and to the best of our
knowledge is not present in the existing literature. The effective
development of electromagnetic theory along the outline traced in
\cite{Gottlieb} is contained in the Hughes lectures
\cite{FeynmanHughes}, where he starts from obtaining Maxwell's
homogeneous equations by writing down a relativistically invariant
action for a charged particle coupled to a $4-$vector potential
(this discussion partially parallels the treatment in
Landau-Lifshitz \cite{Landau:1982dva}). This action is the basis
of an elegant discussion of gauge invariance. After that, he goes
on developing the full Maxwell equations and studying their
consequences. The first two developments were reconstructed and
presented in detail by the present authors in
\cite{DeLuca:2019ija}. The purpose of the present note is to study
the derivation of the non homogeneous Maxwell equations given by
Feynman in \cite{FeynmanHughes}, and to compare his derivation of
the entire set of Maxwell equations (both homogeneous and non
homogeneous), appearing in those lectures, with the ones existing
in the literature. As noted in \cite{DeLuca:2019ija}, unlike his
derivation of the Lorentz force and his discussion of the
homogeneous Maxwell equations and of the related gauge invariance,
the discussion of the inhomogeneous Maxwell equations requires
some more physical input. This is not surprising since -- unlike
the Lorentz force which describes the action of fields on charges,
and the homogeneous Maxwell equations which describe general
constraints on the fields -- the inhomogeneous equations describe
the effect of charges on fields. In fact, from the action
principle point of view, to find these equations one needs an
action functional for the fields themselves, while the Lorentz
force and the homogeneous equations, as already recalled, can be
deduced from the action of the charged particle with the
interaction term of the particles with the field.

Feynman's approach to electromagnetism bears some similarities to
other existing approaches (besides Landau's one) which aim to
derive electromagnetism from relativity. Indeed, from an advanced
point of view, a popular and elegant alternative to introduce
electromagnetism is to make it descend from relativity rather than
the other way round, which is the standard historical route.
Already starting from Page's pioneering paper in 1912 \cite{Page},
it became clear that magnetic fields could be understood as
relativistic effects due to relative motion with respect to
electric charges. In particular, the magnetic field of a uniformly
moving point charge (and in fact, the whole of magnetostatics if
the superposition principle is assumed) can be obtained by
boosting the Coulomb field of a static charge
\cite{Page,Purcell,French, RosserRel1, RosserRel2, Schroeder}.
Also, the fields of accelerating charges, including radiation, can
be derived from the knowledge of the Coulomb field and relativity,
with a few more assumptions, i.e. the action and reaction
principle, charge conservation and the request that the
electromagnetic interaction propagates with the speed of light
\cite{Tessman}. Once known the general fields produced by a moving
charge, the whole set of Maxwell's equations can be derived using
the superposition principle \cite{Page1922,
Page1940,Rosser,Rosser2}. An alternative approach starts from the
fundamental equation of electrostatics, namely Gauss' theorem
(i.e. the first Maxwell equation), which in turn can be derived
from the Coulomb force plus the superposition principle, and to
boost it \cite{Corson}, or to ``covariantize'' it \cite{Schwartz}.
More sophisticated and abstract approaches exist as well
\cite{Landau:1982dva, Susskind3}, in which the superposition
principle is assumed while the Coulomb law is not. Rather, an
action for the electromagnetic field is constructed by assuming
the latter principle, plus relativistic and gauge
invariance\footnote{The latter is a consequence of the homogeneous
Maxwell equations, which can be obtained independently.}, whose
variation gives the non homogeneous Maxwell equations. In such an
approach, the Coulomb law is derived as the solution of the
equations for a point static source. 
%

In \cite{FeynmanHughes}, after the discussion of relativity, the
Lorentz force and the derivation of the homogeneous equations,
Feynman goes on to complete electromagnetism, and in fact he
adopts an approach which is analogous to the above cited ones.
However, despite the analogies, Feynman's approach is quite
original. Consistently with his approach to the homogeneous
equations, Feynman chooses to work with the potentials -- rather
than the fields -- from the beginning. This is certainly due to
the simpler mathematics involved, but also to Feynman's experience
in quantum theories, where potentials are not just a convenient
mathematical tool, but rather a necessary ingredient for the
formulation of the theory, acquiring a physical reality on their
own. In this respect, Feynman's approach is, to the best of our
knowledge, different from all the ones existing in the literature,
and shows the customary originality that is proper of Feynman's
work. Hence, it may be worth to describe Feynman's derivation of
the non homogeneous Maxwell equations in detail. This article
gives then a complete and original formulation of Maxwell's
equations, fully consistent with the outline suggested by Feynman
in the unpublished notes \cite{Gottlieb}.

We complete this work by providing some considerations concerning
the foundations of special relativity, which are naturally
suggested by any approach aiming to derive electromagnetism from
relativity, although this is not addressed at all in Feynman's
notes. In particular, we discuss how the principles and results of
special relativity can be obtained without reference to
electromagnetism, which is necessary for the logical consistency
of this kind of approach to the latter. If special relativity is
treated in this way, the privileged speed which is invariant and
acts as a universal speed limit is left undetermined. It is then
the request of covariance for the laws of electromagnetism which
fixes it to be equal to the speed of electromagnetic waves.

The paper is organized as follows. In Sect. \ref{Hom} we briefly
recall the derivation of the homogeneous Maxwell equations, and of
the related gauge invariance, which we first presented in
\cite{DeLuca:2019ija}, emphasizing the similarities and the
differences with the Landau-Lifshitz approach
\cite{Landau:1982dva}. Related to this, we attempt to understand
whether Feynman took this approach from Landau's book, or
developed it independently. In Sect. \ref{NonHom}, the derivation
of the nonhomogeneous equations given by Feynman in
\cite{FeynmanHughes} is described in detail. In Sect.
\ref{Potentials}, we discuss Feynman's choice of giving priority
to the potentials. Finally, after the concluding section, in the
appendix, we briefly review the discussion of special relativity
without electromagnetism.

\section{The derivation of the homogeneous equations and gauge invariance}
\label{Hom}

Before getting to the heart of the present discussion, we here
summarize Feynman's derivation of the homogeneous Maxwell
equations, and of the associated gauge invariance, as developed in
\cite{FeynmanHughes}, and discussed in \cite{DeLuca:2019ija} (to
which we refer for details).

The homogeneous Maxwell equations emerge naturally in the search
of a generalization of the least action principle to the
relativistic case. In this case, of course the action has to be a
Lorentz scalar, in order for equations of motions to be covariant.
A first candidate for relativistic equation of motion for a point
particle with mass $m_0$ could be:
\begin{eqnarray}\label{RelEOM1}
\frac{\rm d}{{\rm
d}t}\left[\frac{m_0\mathbf{v}}{\sqrt{1-\frac{v^2}{c^2}}}\right]=-\nabla
{\cal V},
\end{eqnarray}
which can be obtained from the action
\begin{eqnarray}\label{RelAct1}
S=\int_{t_1}^{t_2}\left[-m_0c^2\sqrt{1-\frac{v^2}{c^2}}-{\cal
V}(\mathbf{x},t)\right] {\rm d}t = \int_{t_1}^{t_2} [-m_0 c\, {\rm
d}s - {\cal V}(\mathbf{x},t) \, {\rm d}t] \,
\end{eqnarray}
(where ${\rm d}s=\sqrt{1-v^2/c^2} \, c{\rm d}t$), However, this is
not a good choice since the potential {energy} term $ {\cal
V}(\mathbf{x},t)\, {\rm d}t$ is evidently not invariant. A
straightforward modification would be replacing this term with an
invariant scalar potential term $\mathcal{X}(x,y,z,t) \, {\rm d}s$
but, as quickly stated by Feynman (see also the discussion in
Sect. \ref{Potentials}), such a term does not lead to any known
law of nature. The next simplest possibility is then an invariant
term of the form $A_{\mu}(x,y,z,t)dx^{\mu}$, where $A^{\mu}$ is a
$4-$vector, so that the action is
\begin{eqnarray}\label{RelAct2}
S = \int_{t_1}^{t_2} [-m_0 c \, {\rm d}s - A_{\mu} \, {\rm
d}x^{\mu}] \, .
\end{eqnarray}
In the following we shall set $A^{\mu}=(\phi/c,\mathbf{A}$), thus
matching the usual notation. This action becomes stationary when
\begin{eqnarray}\label{RelEOM2}
\frac{\rm d}{{\rm
d}t}\left[\frac{m_0\mathbf{v}}{\sqrt{1-\frac{v^2}{c^2}}}\right]=
\mathbf{F},
\end{eqnarray}
with the force $\mathbf{F}$ given by:
\begin{eqnarray}\label{EoM}
\mathbf{F}=-\nabla\phi-\frac{\partial}{\partial t}\mathbf{A} +
\mathbf{v}\times(\nabla \times \mathbf{A}).
\end{eqnarray}
This expression has the same structure as the Lorentz force acting
on a particle of charge $q$, upon the redefinition
$A_{\mu}\rightarrow qA_{\mu}$, and a direct comparison leads to
the identification
\begin{eqnarray}\label{fields}
\mathbf{E}=-\nabla\phi-\frac{\partial}{\partial
t}\mathbf{A};\qquad \mathbf{B}=\nabla\times\mathbf{A} \, ,
\end{eqnarray}
from which we can immediately obtain the homogeneous Maxwell
equations by using standard vector calculus identities:
\begin{eqnarray}\label{HomogeneousMaxwell}
\nabla\cdot\mathbf{B}=0,\qquad \nabla\times\mathbf{E}=
-\frac{\partial}{\partial t}\mathbf{B}.
\end{eqnarray}

This least action framework also allows a nice discussion of gauge
invariance. Consider again the action (\ref{RelAct2}), in the form
\begin{eqnarray}
S=-m_0c\int \! \,{\rm d}s - \int \! (\phi  -
\mathbf{A}\cdot\mathbf{v})\, {\rm d}t \, .
\end{eqnarray}
Gauge invariance emerges from the question of whether it is
possible to change the potentials while keeping the same minimum
of the action, i.e. the same physics. This evidently happens if
the difference of the actions written in terms of two different
sets of potentials
\begin{eqnarray}\label{difference1}
S-S'=-\int \! (\varphi -\mathbf{a}\cdot\mathbf{v}) \, {\rm d}t \,
,
\end{eqnarray}
(where $\varphi=\phi-\phi'$, $\mathbf{a}=\mathbf{A}-\mathbf{A}'$)
is independent of the integration path, which happens if the
integrand is a total differential of some function $\chi$, i.e.
$\varphi -\mathbf{a}\cdot\mathbf{v}={{\rm d}\chi(x,y,z,t)}/{{\rm
d}t}$, so that
\begin{eqnarray}\label{difference2}
S-S'&=&-\int_{t_i}^{t_f} \frac{{\rm d}\chi(x,y,z,t)}{{\rm
d}t}\,dt=-\int_{t_i}^{t_f} \left(\frac{\partial\chi}{\partial t} +
\frac{\partial\chi}{\partial x}\dot{x} +
\frac{\partial\chi}{\partial y}\dot{y}+
\frac{\partial\chi}{\partial z}\dot{z}\right) {\rm d}t \nonumber \\
&=& -\chi(x,y,z,t_f)+\chi(x,y,z,t_i) \, .
\end{eqnarray}
By comparing (\ref{difference1}) with (\ref{difference2}) we see
that such condition is satisfied if $\varphi ={\partial
\chi}/{\partial t}$ and $\mathbf{a}=-\nabla \chi$, which is
equivalent to the gauge transformations
\begin{eqnarray}\label{gauge}
\phi'=\phi - \frac{\partial \chi}{\partial t}\, , \qquad
\mathbf{A}'=\mathbf{A}+\nabla \chi \, .
\end{eqnarray}
The fact that the equations of motion do not change under such
transformations is also evident from the observation that the
electric and magnetic fields as expressed in (\ref{fields}), are
left unchanged.


Attentive readers have surely noticed the similarity between the
above derivation of the homogeneous Maxwell equations and the
treatment given by Landau and Lifshits (see \cite{Landau:1982dva},
Chapt. 3). The similarity is almost complete, the difference lying
mainly in the fact that Landau \emph{defines} the electric force
as the velocity independent part of the force (\ref{EoM}), and the
magnetic force as the corresponding velocity dependent part of
(\ref{EoM}), so that expressions of the electric and magnetic
fields are extracted from this expressions as the coefficients of
$q$ and of $q\mathbf{v}$ respectively, i.e. the parts that do not
depend on the charge which feels the force. On the other hand,
Feynman had already independently obtained the expression of the
Lorentz force, so that he identifies the electric and magnetic
fields after comparison with (\ref{EoM}). Also, Landau's treatment
of gauge invariance goes in the inverse way, by noticing that a
transformation of the $4-$potential of the form (\ref{gauge}) only
results in a total derivative term in the action, instead of
asking what kind of transformations results only in the addition
of a total derivative.

It is then natural to ask whether Feynman took inspiration from
Landau and Lifshits' textbook. This is not at all an obvious
question, since although this book was already available (in an
earlier edition) at the time, Feynman notoriously did not like
reading the existing literature and textbooks, and preferred
developing everything from scratch. Then it is likely that he
independently envisaged this approach to the homogeneous Maxwell
equations, also given the fact that the subsequent derivation of
the inhomogeneous equations (see next section) differs
considerably from Landau and Lifshits' one. Another possibility is
that he only knew about the general idea and rebuilt everything
from it\footnote{It is well known that, when he needed to know the
result of a paper, he preferred to work in this way rather than
fully reading it.}. On the other hand, it is well known that
Feynman had the highest consideration of Landau (see e.g.
\cite{Shifman}), although they never met in person. In fact, given
Feynman's interest (in the 1950s) concerning condensed matter
physics, he knew very well the pioneering work of Landau on the
subject, upon which he developed his treatments of superfluidity,
polarons and superconductivity (see e.g. \cite{Feynman:1957xma}
and \cite{FeynStatMech}). During his work, he verified many of
Landau's predictions, some of which -- in the first place -- he
did not even believe in, thus getting a feel for Landau's skill.
The reverse is also true. Landau especially admired Feynman's
diagram technique -- which he felt he would have not been able to
develop himself -- rating him as first class in his famous
classification of physicist, i.e. higher than himself
\cite{GinzburgBook}\footnote{Some physicists which knew them both,
e.g. V. L. Ginzburg \cite{GinzburgPT, Goodstein}, thought the two
men to be so similar in many ways, that this similitude had to be
genetic in nature (Feynman's father was actually of belarusian
origins).}.

Thus, the possibility that Feynman indeed had studied, or at least
consulted, Landau's book, taking inspiration for his treatment of
homogeneous Maxwell equations, exists. We believe that this issue
is not settled, and it certainly deserves a more accurate
investigation.

\section{The derivation of the inhomogeneous equations}
\label{NonHom}

In the second chapter of \cite{FeynmanHughes}, Feynman embarks on
a somewhat conventional and rather articulate discussion of
electrostatics, which closely follows the one already given in the
Caltech lectures, the starting points being the Coulomb
\emph{potential}, and the principle of superposition. Then
discussions are given about the fields created by various charge
distributions, as well as Gauss' theorem, until deriving the
Poisson equation, which is of course equivalent to the first
inhomogeneous Maxwell equation. After that, condensers (adopting
the variational technique also introduced in the lectures) and
dielectrics are discussed. As typical of the Hughes lectures, some
off-topic remarks are scattered here and there, concerning in
particular the atomic bomb and a possible intuitive explanation of
the exclusion principle\footnote{This was one of the things that
Feynman notoriously failed to achieve (see e.g.
\cite{Feynman:1963uxa}, vol. III), although he later did come out
with an answer involving a belt, reported e.g. in
\cite{FeynmanDirac}.}.

In Chapt. 3, Feynman completely bypasses magnetostatics, leaving
it for homework (he will however briefly fill the gap in the
lectures of the following year \cite{FeynmanHughes2}, which are
the follow up of the presently discussed ones), and just jumps to
electrodynamics, where relativity comes into play. The main
problem, Feynman claims, is to understand the law that
``determines how the fields are produced'' (\cite{FeynmanHughes},
p. 105), by making explicit that his focus will be on the
potentials rather than on the fields, just because they are easier
to calculate. This is in line with the above derivations of the
homogeneous equations and various other statements Feynman did at
different times (see the following section). The only result that
is already at hand is, then, the Poisson equation for the static
electric potential:
\begin{eqnarray}
\nabla^2\phi=-\frac{\rho}{\epsilon_0},
\end{eqnarray}
where $\rho$ is the electric charge density. Besides this, Feynman
recalls what is already known, namely the Lorentz force and the
relations between potentials and fields (which are equivalent to
the homogeneous Maxwell equations):
\begin{eqnarray}
\mathbf{F}&=&q(\mathbf{E}+\mathbf{v}\times\mathbf{B});\\
\mathbf{E}&=&-\nabla\phi-\frac{\partial\mathbf{A}}{\partial t};\label{elec}\\
\mathbf{B}&=&\nabla\times\mathbf{A}.
\end{eqnarray}
Before proceeding, however, he dwells a bit on the kind of
arguments that he will follow by computing the electric field
generated by a moving charge, just by taking the Coulomb potential
solution of the Poisson equation, boosting it\footnote{This part
of the computation is reported also in \cite{Feynman:1963uxa},
Vol. II, Chapt. 25.} (recalling from the previous discussion of
the homogeneous Maxwell equations that $\phi$ and $\mathbf{A}$ are
components of the same $4-$vector $A^{\mu}$), and using Eq.
(\ref{elec}) to get the field. Instead, he does not compute the
magnetic field, limiting himself to a qualitative discussion. As
stated, this is analogous to the approach of
\cite{Page,Purcell,French,RosserRel1,RosserRel2,Schroeder},
although the authors of these references directly compute the
magnetic field, while Feynman, coherently with his declarations,
computes the potentials first, and then derives the fields.

%

Coming back to the original problem, he then proposes to guess the
general law by extending the Poisson equation to a
relativistically covariant one. The simplest thing to do is to add
a second time derivative, so to get the inhomogeneous D'Alembert
equation:
\begin{eqnarray}\label{Dalembert0}
\left(\nabla^2-\frac{1}{c^2}\frac{\partial^2}{\partial
t^2}\right)\phi =-\frac{\rho}{\epsilon_0},
\end{eqnarray}
where the $1/c^2$ factor appears for dimensional reasons, and the
invariance of $c$ ensures that the operator
$\nabla^2-\frac{1}{c^2}\frac{\partial^2}{\partial t^2}$ is
invariant.  Since, as already known, $\phi$ is not a scalar but
the time component of a $4-$vector, this equation has to be
supplemented with analogous ones for the remaining space
components:
\begin{eqnarray}\label{Dalembert1}
\left(\nabla^2-\frac{1}{c^2}\frac{\partial^2}{\partial
t^2}\right)\mathbf{A} =-\frac{\mathbf{j}}{\epsilon_0\,c^2}.
\end{eqnarray}
In Eq. (\ref{Dalembert1}), the tacit statement is made that the
current density $\mathbf{j}$ is the spatial part of a $4-$vector
$j^{\mu}$, which has $c\rho$ as time component, i.e.
$j^{\mu}=(c\rho,\mathbf{j})$ (this is discussed by Feynman shortly
afterwards in a standard way, hence we do not reproduce his
discussion). The four equations (\ref{Dalembert0}) and
(\ref{Dalembert1}) can be written in the manifestly covariant
compact form
\begin{eqnarray}
\Box A^{\mu} =\frac{j^{\mu}}{\epsilon_0\,c^2}.\label{Dalembert}
\end{eqnarray}
where $\Box=\frac{1}{c^2}\frac{\partial^2}{\partial t^2}-\nabla^2$
and, we recall, $A^{\mu}=(\phi/c, \mathbf{A})$.

Of course, Eqs. (\ref{Dalembert}) describe waves propagating in
space with speed $c$. This can be stated in another form, namely
that the request of Lorentz invariance of the wave equation forces
the waves to move with the speed of light\footnote{Notice that by
extending the relativistic covariance of the dynamical equations,
rather than that of their solutions, it is not necessary to assume
that electromagnetic interactions propagate with the speed of
light (as e.g. is done in \cite{Tessman}), since this is included
in the wave equation we wrote.}. Although Feynman does not say
anything about this matter, it is important for us to notice here
that, if we adopt an approach to special relativity such as one of
those described in Appendix \ref{Relativity}, i.e., independent of
any previous knowledge of electromagnetism and hence with an
unidentified invariant speed $V$, the only way for the above
equation to be relativistically covariant is to have that
invariant speed in front of the time variable $t$. Any other
choice would spoil covariance. This means that if electromagnetism
is constructed with the constraint of relativistic invariance,
then the identification of the invariant speed with the speed of
electromagnetic waves is a necessary consequence. This statement
is valid not only for the present case, but for any \emph{ab
initio} relativistic formulation of electromagnetism, such as the
ones presented in
\cite{Tessman,Page1922,Page1940,Rosser,Rosser2,Corson,Schwartz,Susskind3}.

Feynman does not neglect mentioning that Eq. (\ref{Dalembert})
does not specify the $4-$potential $A^{\mu}$ uniquely, since it is
always possible to add to it a $4-$vector $\chi^{\mu}$ satisfying
the homogeneous D'Alembert equation $\Box \chi^{\mu}=0$. From a
standard textbook point of view, this is of course nothing that
the residual gauge invariance after having chosen the Lorenz gauge
condition, which is required in order to get (\ref{Dalembert})
from Maxwell's equations. Feynman deals with this ambiguity by
invoking the physical assumption that no waves without source
exist. It is at this point that he makes the epistemological
remarks that ``there exists an arbitrariness in what you assume
and what you prove'', hence ``one man's assumption is another
man's conclusion'' (\cite{FeynmanHughes}, p. 108), which fit well
with his general attitude to physics and with the presently
described way of dealing with electromagnetism. In the following
section, he writes down an action principle for the potential part
of Eq. (\ref{Dalembert}):
\begin{eqnarray}
S_A=\frac{1}{2}\int
\left(\partial_{\nu}A_{\mu}\right)\left(\partial^{\mu}A^{\nu}\right)\,d^4x
\end{eqnarray}
(which is nothing but the Fermi action for the Lorenz gauge-fixed
electromagnetic field \cite{Fermi:1932xva}). Adding this action to
(\ref{RelAct2}), which describes charged particles and their
interaction with the electromagnetic field, gives the full (in his
words, ``superduper'') action from which the whole of
electrodynamics and mechanics can be obtained.

The above equations are not yet the usual Maxwell equations, of
course, but before moving on to write them down Feynman discusses
one more assumption -- namely causality -- or that cause always
preceeds effects\footnote{This is related to the request that all
waves must have a source.}. We have already discussed elsewhere
\cite{FeynAmplifier} the relevance of the causality issue in
Feynman's approach to physics. Here he uses this assumption to
restrict the general solution to Eq. (\ref{Dalembert}) to the
usual retarded $4-$potential:
\begin{eqnarray}\label{retarded}
A_{\mu}(\mathbf{r}, t)=\frac{1}{4\pi
\epsilon_0\,c^2}\int\frac{j_{\mu}(\mathbf{r}',t-\frac{|\mathbf{r}-\mathbf{r}'|}{c})}{|\mathbf{r}-\mathbf{r}'|}
\,d^3\mathbf{r}'.
\end{eqnarray}
Here Feynman could not resist from including a digression about
his old work with Wheeler on the absorber theory, where time
symmetric solutions to Eq. (\ref{Dalembert}) are studied. The
application to the problem of the self force of charged particles
and of radiation resistance, whose solution was the main thrust
behind the Wheeler-Feynman theory, follows. Also, a brief
discussion of another piece of Feynman's old work on
electrodynamics is discussed, namely the field-free formulation
which played a major role in his Nobel prize winning work on QED.
Although thought-provoking, these remarks are not instrumental to
what follows, so we refer the interested reader to
\cite{FeynmanHughes}.

Finally, Feynman comes back to his original task of writing down
Maxwell's equations, i.e. to ``finishing the above set of
equations'' (\cite{FeynmanHughes}, p. 128). For this he adds one
last physical assumption regarding charge conservation, which is
expressed by the continuity equation
\begin{eqnarray}
\frac{\partial \rho}{\partial t}=-\nabla\cdot\mathbf{j}\,,
\end{eqnarray}
which he uses to find a relation between the derivatives of the
solution (\ref{retarded}), which is\footnote{Observe that this
would be true for any linear combination of retarded and advanced
solutions, that is, for any solution of (\ref{Dalembert}), so that
causality is not a necessary assumption to get the Maxwell
equation (in fact, these equations are time-symmetric).}
\begin{eqnarray}\label{Lorenz}
\frac{1}{c^2}\frac{\partial\phi}{\partial t}=
-\nabla\cdot\mathbf{A}\,.
\end{eqnarray}
This is the Lorenz gauge condition (although Feynman inaccurately
refers to it as a gauge ``transformation''). As already remarked,
in usual approaches to electromagnetism this condition is imposed
to reduce the full Maxwell equations to the wave equation for the
potentials (\ref{Dalembert}). Here instead (\ref{Dalembert}) is
the starting point, so its solutions have to satisfy
(\ref{Lorenz}). Now he has all the ingredients to compute the curl
of $\mathbf{B}$, finding finally the differential
Amp\'{e}re-Maxwell equation:
\begin{eqnarray}
\nabla\times\mathbf{B}=\nabla\times(\nabla\times\mathbf{A})=-\nabla^2\mathbf{A}+\nabla(\nabla\cdot\mathbf{A})=\frac{\mathbf{j}}{\epsilon_0\,c^2}+\frac{1}{c^2}\frac{\partial}{\partial
t} \left(-\frac{\partial\mathbf{A}}{\partial
t}-\nabla\phi\right)=\frac{\mathbf{j}}{\epsilon_0\,c^2}+\frac{1}{c^2}\frac{\partial\mathbf{E}}{\partial
t}.
\end{eqnarray}
Analogously, he recomputes the divergence of $\mathbf{E}$, getting
again the differential form of Gauss' theorem:
\begin{eqnarray}
\nabla\cdot\mathbf{E}=\nabla\cdot\left(-\frac{\partial\mathbf{A}}{\partial
t}-\nabla\phi\right)=\frac{1}{c^2}\frac{\partial^2\phi}{\partial
t^2}-\nabla^2\phi=\frac{\rho}{\epsilon_0}.
\end{eqnarray}
Now the derivation of the full Maxwell equations is complete. We
observe that, although Feynman's procedure, making explicit use of
the potentials, is not gauge invariant, the final result is
instead invariant under the transformations (\ref{gauge}). The end
result is thus:
\begin{eqnarray}
\nabla\cdot\mathbf{E}=\frac{\rho}{\epsilon_0},\qquad
\nabla\times\mathbf{B}=\frac{1}{c^2}\left(\frac{\mathbf{j}}{\epsilon_0}+\frac{\partial\mathbf{E}}{\partial
t}\right).
\end{eqnarray}
After concluding his derivation of the full set of Maxwell
equations, Feynman can move on to the description of their
solutions and applications, which constitute the remaining part of
\cite{FeynmanHughes}. But for us now it is time to stop and
ponder.

\section{Feynman and the importance of electromagnetic potentials}
\label{Potentials}

As already repeatedly stated above, the main difference with other
\emph{ab initio} relativistic approaches to electromagnetism
present in the literature is the emphasis that Feynman gives to
potentials. This emphasis is of course dictated by Feynman's
experience in quantum mechanics and the least action principle.
Quite illuminating, in this respect, is the following quotation
\cite{FeynMercereau}:
\begin{quote}
Yet, the Schr\"odinger equation can only be written neatly with
$\bf A$ and $V$\footnote{This $V$ corresponds of course to what we
are calling $\phi$ in this paper.} explicitly there and it was
pointed out by Bohm and Aharonov (or something like that), that
this means that the vector potential has a reality and that in
quantum mechanical interference experiments there can be
situations in which classically there would be no expected
influence whatever. But nevertheless there is an influence. Is it
action at a distance? No, $\bf A$ is as real as $\bf B$-realer,
whatever that means.
\end{quote}
A similar view was repeatedly expressed also in the Caltech
lectures \cite{Feynman:1963uxa}, for example in Sect. 15-5 of
volume II he states that:
\begin{quote}
What we mean here by a ``real'' field is this: a real field is a
mathematical function we use for avoiding the idea of action at a
distance... A ``real'' field is then a set of numbers we specify
in such a way that what happens at a point depends only on the
numbers at that point... In our sense then, the $\mathbf{A}$-field
is ``real'' [...] the vector potential $\mathbf{A}$ (together with
the scalar potential $\mathbf{\phi}$ that goes with it) appears to
give the most direct description of the physics. This becomes more
and more apparent the more deeply we go into the quantum theory.
In the general theory of quantum electrodynamics, one takes the
vector and scalar potentials as the fundamental quantities in a
set of equations that replace the Maxwell equations: $\mathbf{E}$
and $\mathbf{B}$ are slowly disappearing from the modern
expression of physical laws; they are being replaced by
$\mathbf{A}$ and $\mathbf{\phi}$.
\end{quote}

For Feynman, indeed, the foundations of electromagnetism are
essentially \emph{quantum}\footnote{Such a view has been expressed
also by other authors, for example \cite{Umezawa:1982nv}, and
especially \cite{Mead}, who explicitly acknowledges having been
deeply inspired by Feynman.}, as also expressed in the Hughes
lectures, where he states that (see \cite{FeynmanHughes}, p. 35):
\begin{quote}
[...] forces which we shall call fundamental or primary which are
conservative. [...] I shall call conservative forces, those forces
which can be deduced from quantum mechanics in the classical
limit.
\end{quote}
This is, evidently, a remarkable point for using potentials even
in classical electromagnetism.  According to M. A. Gottlieb,
quoting Matthew Sands, Feynman expressly said that ``he would
start with the vector and scalar potentials, then everything would
be much simpler and more transparent'' \cite{StackExchange}.

A further possible motivation for Feynman to stress the role of
the potentials can be found in his work in quantum field theory.
In his lectures on gravitation (\cite{Feynman:1996kb}, ch. 3), as
well as in his Hughes lectures on astronomy and astrophysics
(\cite{FeynmanHughes1}, pp. 30-31), he embarks on a careful
discussion about the properties of the forces mediated by fields
with different spin. Of course, for that discussion, his aim was
to show that gravitation has to be described by a spin-2 field
(the graviton), in order to match what is known about general
relativity. However, he also states explicitly what is the
behavior of the charges associated with different spin fields
under Lorentz boosts. In particular, the scalar charge would
decrease (this was the \emph{ratio} under his rather cryptic
statement in \cite{FeynmanHughes} p. 42, also recalled in Sect.
\ref{Hom}, that the action of a relativistic particle in a scalar
4-potential did not give rise to known physics), while the tensor
charge would rather grow, as appropriate for energy-momentum,
which is the source of gravity. The charge associated with a
spin-1 field, instead, is invariant, as the electric charge must
be (obviously neglecting charge renormalization effects, which are
however relevant only at very short scales). Hence
electromagnetism must be described by a spin-1 field, that is, by
a vector potential. In his derivation of the Lorentz force
\cite{DeLuca:2019ija,Gottlieb}, Feynman starts precisely from
these considerations. Since the graduate course on which Ref.
\cite{Feynman:1996kb} is based was given in 1962/63, i.e.
immediately before this derivation was sketched (the date on the
notes \cite{Gottlieb} is December 1963), it is not unlikely that
Feynman's thougths on gravity were the basis of his derivation of
the Lorentz force and of his reformulation of classical
electromagnetism.

Interestingly, the statement of the Lorentz invariance of the
source of a spin-1 field is enough for him, and apparently no
mention is made of the equally crucial fact that only a spin-1
field has the property of generating a repulsive force between
equally charged fields\footnote{When excluding, of course, fields
of higher odd spin, for which however it is not known how to build
a consistent Lorentzian interacting quantum theory.} (this is of
course the reason that is usually given in the literature).

\section{Conclusions}

In the present work, we have described Feynman's reformulation of
the Maxwell equations, along the scheme which was outlined by
Feynman himself in \cite{Gottlieb} and carried out in
\cite{FeynmanHughes}. By a detailed comparison with many other
existing approaches to electromagnetism, we showed that Feynman's
one is unique in the fact that it states the primacy of potentials
with respect to fields, in this reflecting his views on the
quantum foundations of the fundamental interactions, as explicitly
declared in \cite{FeynmanHughes}, but also in several other
places. Also, unlike most other approaches, the present one
clearly singles out the physical hypotheses needed to state the
homogeneous and the non homogeneous Maxwell equations, thus
underlining the crucial difference between the two sets.
Concerning the homogeneous equations, the apparent similarity of
Feynman's approach with Landau's one prompted us to add some
reflections on the relationship between the two, aiming to
understand whether Feynman took inspiration from Landau. About
this, no definite conclusion could be reached, since a much more
thorough and complete investigation of the sources is needed.

Although no hint at this is present in any of Feynman's writings,
we believe that a fully \emph{ab initio} approach to
electromagnetism can only be logically consistent (however this is
by no means necessary for didactic purposes) if special relativity
is developed in a way that is not dependent on electromagnetic
concepts. This way of introducing special relativity is not new,
but has never been part of the mainstream of physics, and to the
best of our knowledge has never been discussed in connection with
such formulations of electromagnetism. For the reader's
convenience we briefly review this matter in the appendix, where
references to the literature are also given.

The work described here can be rightfully considered a new piece
of Feynman's multifaceted jigsaw, reflecting his continuing search
of originality and bearing deep links with other parts of his
scientific production of those years. We also believe that it
shows that, despite a lot has already been said about Feynman's
physics, many more gems are awaiting to be discovered in what he
left unpublished.


\section*{Acknowledgements}

M.D. and A.N. would like to thank M. Consoli and P. Rossi for
suggesting us to consider of the relation between Feynman's and
Landau's work.

\begin{appendices}

\numberwithin{equation}{section}

\section{Some considerations on special relativity without electromagnetism} \label{Relativity}

In order to make electromagnetism descend from special relativity,
logical consistency requires to develop the latter without
reference to the former. This is not usually pointed out in
references which try to develop electromagnetism in this way. It
is the purpose of this appendix to briefly review how this can be
done. This view is in contrast with the usual way in which special
relativity is addressed in textbooks, which start from Einstein's
two postulates: the validity of the principle of relativity for
all physical phenomena in inertial frames, and the invariance of
the speed of light under the transformations that implement the
principle of relativity. Such an approach is of course dependent
on the speed of light, which is an electromagnetic quantity, and
reflects the historical development of Lorentz transformations and
relativity, which famously emerged from the problems and
contradictions of the electrodynamics of moving bodies. As well
known, in fact, Lorentz derived his transformations to explain the
negative result of the Michelson-Morley experiment, while Einstein
was driven by the desire to make electromagnetism fully compatible
with the principle of relativity (see e.g. \cite{Pais} and
references therein).

After a few years since Einstein's breakthrough paper
\cite{Einstein:1905ve}, it was realized by some people
\cite{Ignatowski,FrankRothe,Pars} that special relativity could be
freed from the second postulate, which could be replaced by some
natural and more general assumptions, independent of light and
electromagnetism. In fact, these authors showed that the
requirement of relativity of inertial frames, together with
natural assumptions of homogeneity of space and time, isotropy of
space, and group structure (the group character of the Lorentz
transformations was established by Poincar\'e in 1906
\cite{Poincare}), led to the recognition that the most general
transformations between inertial frames constitute a one parameter
family. In the simplest case of two reference frames with parallel
axes, with relative velocity oriented along the $x$ and $x'$ axes,
such transformations look like:
\begin{eqnarray}\label{Lorentz}
x'&=&\frac{x-vt}{\sqrt{1-\alpha v^2}}\\\nonumber
t'&=&\frac{t-\alpha v x}{\sqrt{1-\alpha v^2}}.
\end{eqnarray}
It is possible to argue on physical grounds that the parameter
$\alpha$, which has the dimensions of the inverse square of a
velocity, has to be non negative. The usual argument (see e.g.
\cite{LevyLeblond}) is that this is required in order to have
causality, i.e. to ensure that there exists at least a class of
events such that the sign of the spacetime interval between pairs
of them is invariant, so that their time ordering is preserved.
Being non negative, this parameter can be actually seen as the
inverse square of a quantity $V$ with the dimensions of speed,
which can also be infinite, i.e. $\alpha=\frac{1}{V^2}$. When one
does so the transformations look formally identical to Lorentz
transformations, with $V$ in place of $c$, while they reduce to
Galilei transformations\footnote{This name was given to them by
Philipp Frank, one of the discoverers of this approach, ``Um einen
kurzen Namen zu haben'' (``in order to have a short name'',
\cite{Frank}, p. 897).} when $V=\infty$, and absolute causality is
recovered. One then sees in the standard way that the
transformations (\ref{Lorentz}) leave $V$ invariant, that they
make no sense for relative velocities greater than this speed, and
that the law of composition of velocities that follows from them
is such that the result can never exceed $V$. Going on, when one
considers a point particle dynamics which is covariant under these
transformations, one finds out that no massive particle can be
accelerated to a greater speed, while any massless particle (i.e.
not just photons) constantly move at that speed. All this of
course implies that no signal can propagate with a speed greater
than $V$. The value of $V$ determines the class of events whose
time ordering is preserved under (\ref{Lorentz}), i.e. the events
that can be connected by a signal, for which $V^2\Delta t^2 \geq
|\Delta \mathbf{x}^2|$.

This approach was later forgotten, probably because of the
skepticism expressed by Pauli in his famous encyclopedia article,
in which he declares he does not like the fact that the invariant
speed is not immediately identified with the speed of
light\footnote{``Nothing can naturally be said about the sign,
magnitude, and physical meaning of $\alpha$. From the
group-theoretical assumption it is only possible to derive the
general form of the transformation formulae, but not their
physical content'', (\cite{Pauli}, p.11). However, Pauli was
apparently not aware that that the sign of the parameter could be
fixed by the causality requirement, which was shown by later
authors.}, and it was later rediscovered by several authors, who
often proposed it as an alternative way to teach special
relativity (a partial list, besides \cite{LevyLeblond}, includes
\cite{Schwartz1, Terletskii, LeeKalotas, Srivastava, Sardelis,
Mermin, Liberati:2001sd, Pelissetto:2015bva}).

It is to be admitted that such an approach, being based on less
assumptions, naturally leads to an increased analytical
complexity, which probably explains why it has never become
popular in teaching. Nevertheless -- strictly speaking -- while
developing the foundations of electromagnetism from a relativistic
point of view, it is necessary to adopt it. A simpler alternative
exists, however, which is described e.g. in the textbook by Ugarov
\cite{Ugarov}. Here, special relativity is developed starting from
the postulate that a limiting speed exists for any
interaction\footnote{This was also assumed by Einstein when he
critically reexamined the clock synchronization procedure
\cite{Einstein:1905ve}.}, after a critique of the Newtonian
concept of instantaneous interaction (interestingly, Feynman also
expresses some doubts about this concept, in connection with the
Coulomb law, in \cite{Gottlieb}). The existence and value of such
a speed would be fundamental laws of nature, therefore
compatibility with the principle of relativity requires its value
to be an invariant, since otherwise its measurement would allow
the observer to discriminate between different inertial frames.
Also, one could make it arbitrarily large by performing a boost.
From the assumption of the invariance of this speed, and the
principle of relativity, then one can derive Lorentz
transformations in the usual way, with the invariant speed
appearing as a parameter.

In all these approaches, the value of the invariant speed is left
undetermined, so that its identification must come from the
experiments. For example, one can invoke the result of the
Michelson-Morley experiment to identify the invariant speed with
the speed of light. Also, one can use accelerator experiments with
elementary particles to show that there is a limiting speed, no
matter how high the energy. Another possibility, which we discuss
more in detail in Sect. \ref{NonHom}, is that when one develops
electromagnetism requiring compatibility with special relativity,
a necessary consequence is that electromagnetic waves must
propagate with the invariant speed if the theory is to be
covariant. This means that one can use measurements on
electromagnetic waves to fix the value of that speed.



\end{appendices}

\end{document}